\documentclass[twocolumn,showpacs,prb]{revtex4}
\usepackage{graphicx}
\usepackage{dcolumn}
\usepackage{amsmath}
\usepackage{amssymb}

\makeatletter
\def\btt#1{\texttt{\@backslashchar#1}}
\DeclareRobustCommand\bblash{\btt{\@backslashchar}} \makeatother

\renewcommand{\Re}{\mathop\mathrm{Re}\nolimits}
\renewcommand{\Im}{\mathop\mathrm{Im}\nolimits}

\begin{document}

\title{Anomalous Surface Impedance in a Normal-metal/Superconductor Junction
with a Spin-active Interface}

\author{Yasuhiro Asano, Masahiro Ozaki, and Tetsuro Habe }
\affiliation{Department of Applied Physics and Center for Topological Science \& Technology,
Hokkaido University, Sapporo 060-8628, Japan}

\author{Alexander A. Golubov}
\affiliation{Faculty of Science and Technology and MESA+ Institute of Nanotechnology,
University of Twente, 7500 AE, Enschede, The Netherlands}


\author{Yukio Tanaka}
\affiliation{Department of Applied Physics, Nagoya University, Nagoya 464-8603, Japan}

\date{\today}

\begin{abstract}
We discuss the surface impedance ($Z=R-iX$) of a normal-metal/superconductor proximity
structure taking into account the spin-dependent potential at the junction interface.
Because of the spin mixing transport at the
interface, odd-frequency spin-triplet $s$-wave Cooper pairs penetrate into
the normal metal and cause the anomalous response to electromagnetic fields.
At low temperature, the local impedance at a surface of the normal metal shows
the nonmonotonic temperature dependence and the anomalous relation $R>X$.
We also discuss a possibility of observing such anomalous impedance in experiments.
\end{abstract}


\maketitle

\section{introduction}

Physics of odd-frequency Cooper pairs\cite{berezinskii} has been a hot issue since
a theoretical paper pointed out
the existence of odd-frequency pairs in realistic proximity structures~\cite{bergeret}.
There are mainly two ways to create the odd-frequency Cooper pairs in proximity structures.
At first, spin-mixing due to spin-dependent potential should
generate odd-frequency pairs. The authors of Ref.~\onlinecite{bergeret}
considered a ferromagnet / metallic-superconductor junction, where the direction of
magnetic moment near the
interface is spatially inhomogeneous. The spin-flip scattering in such magnetically
inhomogeneous segment
produces the odd-frequency spin-triplet $s$-wave Cooper pairs in the ferromagnet.
This prediction has promoted a number of theoretical
studies~\cite{ya07sfs,braude,eschrig,linder,fominov1,
yokoyama0,yokoyama,halterman}.
Manifestations of triplet pairs were recently observed experimentally as a long-range
Josephson coupling across ferromagnets~\cite{Keizer,Anwar,Robinson,Khaire}.
Alternatively, the odd-frequency pair was suggested in proximity structures
involving a normal metal attached to an odd-parity spin-triplet superconductor
that belongs to the conventional even-frequency symmetry class.
The parity-mixing due to inhomogeneity produces the odd-frequency pairs
even in this case~\cite{tanaka07e}.
The unusual properties of spin-triplet superconducting junctions
due to odd-frequency pairs~\cite{tanaka07L,yt04,yt05r,ya06,ya07,fominov2} were predicted
theoretically.
 Unfortunately, however, we have never had clear scientific evidences
 of odd-frequency pairs in experiments.
This is because physical values focused in experiments have
only indirect information of the frequency symmetry.

In a previous paper~\cite{ya11}, we showed that the surface impedance directly
reflects the frequency symmetry of Cooper pair.
Surface impedance $Z=R-iX$ represents the dynamic response of Cooper pairs
to low frequency electromagnetic field~\cite{mattice,nam}.
The surface resistance, $R$, corresponds to resistance due to normal electrons.
The reactance, $X$, represents power loss of electromagnetic field due to Cooper pairs.
In conventional even-frequency superconductors,
the positive amplitude of the Cooper pair density guarantees a robust relation $R\ll X$
at low temperatures and at low frequencies.
The validity of the relation $R<X$, however, is questionable for odd-frequency
Cooper pairs because the odd-frequency symmetry and \textit{negative pair density}
are inseparable from each other according to the standard theory of superconductivity~\cite{agd}.
We have considered a a normal metal/superconductor (NS) junction where superconductor
belongs to spin-triplet odd-parity symmetry.
We have theoretically shown
that the odd-frequency Cooper pairs in the normal metal lead to
the unusual relationship $R>X$. Therefore observing the relation $R>X$ in experiments can be
a very clear and direct evidence which suggests the existence of odd-frequency Cooper pairs.
 Although the detection of such unusual relation
is possible these days, the fabrication of a well characterized NS junction
using chiral $p$-wave spin-triplet superconductor 
Sr$_2$RuO$_4$~\cite{maeno} is not easy task.
Thus we need to discuss a possibility for observing the unusual relationship $R>X$ in
another accessible proximity structures.

In this paper, we discuss the surface impedance in NS junction consisting of
 a metallic superconductor where pairing symmetry belongs to 
spin-singlet $s$-wave. 
At the junction interface, we introduce a thin ferromagnetic
layer which produces the odd-frequency
spin-triplet $s$-wave Cooper pairs in the normal metal.
The local complex conductivity is calculated based on the linear response theory
using the quasiclassical Green function method. We will conclude that the local impedance
in the normal metal show the unusual relation $R>X$ when the odd-frequency pairs is
dominant in the normal metal.
We also discuss a possibility to detect of the relation in experiments.

This paper is organized as follows.
In Sec.~II, we explain the theoretical model of a NS junction and
the formula
for complex conductivity. The calculated results of
impedance in NS junctions are shown in Sec.~III.
The conclusion is given in Sec.~IV.

\section{Model and Method}
\begin{figure}[ht]
\begin{center}
\includegraphics[width=8cm]{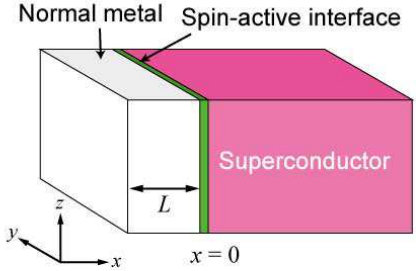}
\end{center}
\caption{(color online). The schematic picture of a normal-metal/superconductor
junction under consideration.
}\label{fig1}
\end{figure}

Let us consider a bilayer of a superconductor and a thin normal metal film
as shown in Fig.~\ref{fig1}, where $L$ is the thickness of the normal metal.
To calculate the complex conductivity in the normal metal, we first solve the quasiclassical Usadel
equation~\cite{usadel} in the standard $\theta$-parameterization,
\begin{equation}
\hbar D \frac{d^2 \theta_\nu(x,\epsilon)}{dx^2} +
2i \epsilon \sin \theta_\nu(x,\epsilon)=0,\label{usadel}
\end{equation}
where $D$ is the diffusion constant of the normal metal and
$\epsilon$ is the quasiparticle energy measured from the Fermi level.
The subscript $\nu=\pm 1$ describes two Nambu spaces: $\nu=1$ indicates
the subspace for electron spin-up and hole spin-down, and $\nu=-1$ indicates
that for electron spin-down and hole spin-up.
Effects of spin-dependent scatterings are considered through the boundary condition
at the NS interface~\cite{kupriyanov,daniel,cottet,linder,yoshizaki},
\begin{align}
\gamma_B \left.\frac{d\theta_\nu}{dx}\right|_{x=0}
=\sin(\theta_\nu-\theta_S) + i\nu \frac{G_\phi}{G_T} \sin\theta_\nu, \label{kl}
\end{align}
where $\gamma_B=L\frac{R_b}{R_d} $ is a interface parameter with $R_d$ and $R_b$
being the resistance of the normal metal and that of the NS interface, respectively.
The Green function in superconductor is described by
\begin{align}
g_S=\cos\theta_S=&\frac{\epsilon}{\sqrt{(\epsilon+i\lambda)^2 -\Delta}}, \label{gs}\\
f_S=\sin\theta_S=&\frac{i\Delta}{\sqrt{(\epsilon+i\lambda)^2 -\Delta}}, \label{fs}
\end{align}
where $\Delta$ is the amplitude of pair potential in the bulk superconductor
and $\lambda$ is a small parameter providing the retarded Green function.
The second term in Eq.~(\ref{kl}) describes the spin-mixing effect at the junction interface.
 $G_T$ represents the spin-independent tunneling conductance of the junction interface,
whereas $G_\phi$ is the spin-mixing conductance~\cite{yokoyama3}.
At the outer surface of the normal metal, we require
\begin{align}
\left. \frac{\partial \theta(x,\epsilon)}{\partial x} \right|_{x=-L}=0.
\end{align}
The normal and anomalous retarded Green functions
are obtained as
\begin{align}
g_\nu(x,\epsilon)=\cos\theta_\nu(x,\epsilon),\;
f_\nu(x,\epsilon)=\sin\theta_\nu(x,\epsilon),
\end{align}
respectively.

Having found the Green functions, we can calculate the local complex conductivity
that describes the response of the sample to the electromagnetic field.
The local complex conductivity
$\sigma_{\textrm{N}}(x,\omega) = \sigma_1+i \sigma_2$ at frequency $\omega$
is determined by the general expression~\cite{FGH}
\begin{align}
\frac{\sigma_1(x,\omega)}{\sigma_0} &= \frac{1}{2\hbar\omega}
 \int_{-\infty}^{\infty} \!\!\!\!\!\! d\epsilon \left[J(\epsilon+\hbar\omega)-J(\epsilon)
\right] K_1,\label{sig1y}\\
\frac{\sigma_2(x,\omega)}{\sigma_0} &= \frac{1}{2\hbar\omega}
 \int_{-\infty}^{\infty}\!\!\!\!\!\! d\epsilon \left[
J(\epsilon+\hbar\omega) K_2  + J(\epsilon) K_3 \right],\label{sig2y}
\end{align}
\begin{align}
K_1 = \sum_{\nu}
f_{\nu,I}&(\epsilon)f_{\nu,I}(\epsilon+\hbar\omega)\! +\!
g_{\nu,R}(\epsilon)g_{\nu,R}(\epsilon+\hbar\omega),\label{k1}\\
K_2 =\! \sum_{\nu} f_{\nu,R}&(\epsilon)f_{\nu,I}(\epsilon+\hbar\omega)\! -\!
g_{\nu,I}(\epsilon)g_{\nu,R}(\epsilon+\hbar\omega),\label{k2}\\
K_3 =\! \sum_{\nu} f_{\nu,R}&(\epsilon+\hbar\omega)f_{\nu,I}(\epsilon)\! -\!
g_{\nu,I}(\epsilon+\hbar\omega)g_{\nu,R}(\epsilon),\label{k3}
\end{align}
with $J(\epsilon)= \tanh\left({\epsilon}/{2k_B T}\right)$ and
\begin{align}
g_{\nu,R}(\epsilon) =& \Re \left[ g_\nu(x,\epsilon) \right],\;
g_{\nu,I}(\epsilon) =
\Im \left[ g_\nu(x,\epsilon) \right], \\
f_{\nu,R}(\epsilon) =& \Re \left[ f_\nu(x,\epsilon) \right], \;
f_{\nu,I}(\epsilon) = \Im \left[ f_\nu(x,\epsilon) \right]. \label{fi}
\end{align}
The local impedance in the normal metal is calculated
from the complex conductivity as
\begin{equation}
Z_{\textrm{N}}(x,\omega)=R_{\textrm{N}}-iX_{\textrm{N}}= (1-i) \sqrt{\frac{\hbar\omega}{\Delta_0}
\frac{\sigma_0}{\sigma_{\textrm{N}}(x,\omega)}}{Z_\textrm{0}},\label{zns}
\end{equation}
where $Z_{\textrm{0}}\equiv \sqrt{2\pi \Delta_0/\sigma_0c^2\hbar}$,
$\Delta_0$ is the amplitude of pair potential at $T=0$, and $\sigma_0$
is the Drude conductivity in the normal metal.
In this paper, we describe the dependence of $\Delta$ on temperature by the BCS theory.
In particular, we focus on the local impedance at the surface of the normal metal
defined by
\begin{align}
R_L-iX_L\equiv Z_{\textrm{N}}(-L,\omega).\label{zl}
\end{align}
Such local impedance is an accessible observable these days~\cite{machida}.
Usual experiments measure the impedance of the whole NS structure which is calculated
as
\begin{align}
Z_{\textrm{NS}}=&R_{\textrm{NS}}\!-\!iX_{\textrm{NS}}=\bar{Z}_{\textrm{N}} \frac{Z_{\textrm{S}} \cos \bar{k}_n L\!
-\! i \bar{Z}_{\textrm{N}} \sin \bar{k}_n L }
{ \bar{Z}_{\textrm{N}} \cos \bar{k}_nL\! -\! iZ_{\textrm{S}} \sin \bar{k}_nL },\label{zns1}
\end{align}
where $Z_S$ is the impedance of superconductor which is obtained by substituting the
Green function of superconductor in Eqs.~(\ref{gs})-(\ref{fs}) into Eqs.~(\ref{sig1y})-(\ref{fi}).
In this paper,  $L$ is chosen to be comparable to $\xi_{T_c}=\sqrt{\hbar D/2\pi T_C}$
with $T_C$ is the superconducting transition temperature.
In such junctions, the conductivity is almost independent of $x$ in the normal metal.
Therefore it is possible to define spatially averaged values
of the conductivity, the impedance and the wavenumber of
electromagnetic field as follows
\begin{align}
\bar{\sigma}_{\textrm{N}} =&  \int_{-L}^{0} dx \sigma_{\textrm{N}}(x)/L, \\
\bar{Z}_{\textrm{N}}=&\bar{R}_{\textrm{N}}-i\bar{X}_{\textrm{N}}
= -i \sqrt{ 4\pi i \omega / (c^2 \bar{\sigma}_{\textrm{N}})},\label{zn}\\
\bar{k}_n=&  \sqrt{ i {4\pi \omega \bar{\sigma}_{\textrm{N}}}/{c^2}  }.
\end{align}

To understand the relation between the frequency symmetry of a Cooper pair and
the sign of the imaginary part of complex conductivity $\sigma_2$,
we analyze the spectral pair density defined by
\begin{align}
K_s(\epsilon) =&\sum_{\nu} f_{\nu,R}(\epsilon)f_{\nu,I}(\epsilon)-
g_{\nu,R}(\epsilon)g_{\nu,I}(\epsilon),\\
=&\sum_{\nu} 2f_{\nu,R}(\epsilon)f_{\nu,I}(\epsilon)
=\sum_{\nu} \textrm{Im}f_\nu^2(\epsilon),\label{ks}
\end{align}
which appears in the integrand of $\sigma_2$ in Eq.\ (\ref{sig2y})
at very small $\omega$. We used the normalization
condition $g_\nu^2(\epsilon)+f_\nu^2(\epsilon)=1$.
The spectral pair density contains full information about
the symmetry of $f(\epsilon)$ and, therefore, the frequency symmetry of Cooper pairs.
At $T=0$, the Cooper pair density in the normal metal is
\begin{align}
n_s =\int_{-\infty}^\infty d\epsilon J(\epsilon) K_s(\epsilon).
\end{align}
Since $K_s$ is an odd function of $\epsilon$ according to its definition
and $J(\epsilon)$ is also odd step function of $\epsilon$, the pair density
becomes
\begin{align}
n_s =2 \int_{0}^\infty d\epsilon K_s(\epsilon). \label{ns}
\end{align}
Finally the local density of states is given by
\begin{align}
N(\epsilon,x)=\sum_{\nu} \textrm{Re}[ g_{\nu}(\epsilon,x)],
\end{align}
which is normalized to the normal density of states at the Fermi level.

\section{Results}
The theory includes several independent parameters discussed as follows.
Throughout this paper, we fix the thickness of a normal metal $L$ at $\xi_{T_c}$.
The spatial dependence of the Green function in the normal metal becomes weak
in this choice. In numerical simulation, we do not discuss details of the
averaged impedance $\bar{Z}_{\text{N}}$ in Eq.~(\ref{zn}) because we have confirmed
that $Z_{L} \approx \bar{Z}_{\text{N}}$.
The second parameter $R_d/R_b$ tunes the degree of the proximity effect in a
normal metal. The larger $R_d/R_b$ gives the stronger proximity effect.
The third one is $G_\phi/G_T$ which represents the strength of spin-dependent
potential at the NS interface.
The forth one is the frequency of electromagnetic field $\omega$ which should be
smaller than $\Delta_0/\hbar$ to obtain information about Cooper pairs.
Finally we fix the small imaginary
part in energy as $\lambda = 0.001 \Delta_0$, which does not affect following conclusions.

\subsection{Density of States}

\begin{figure}[ht]
\begin{center}
\includegraphics[width=8cm]{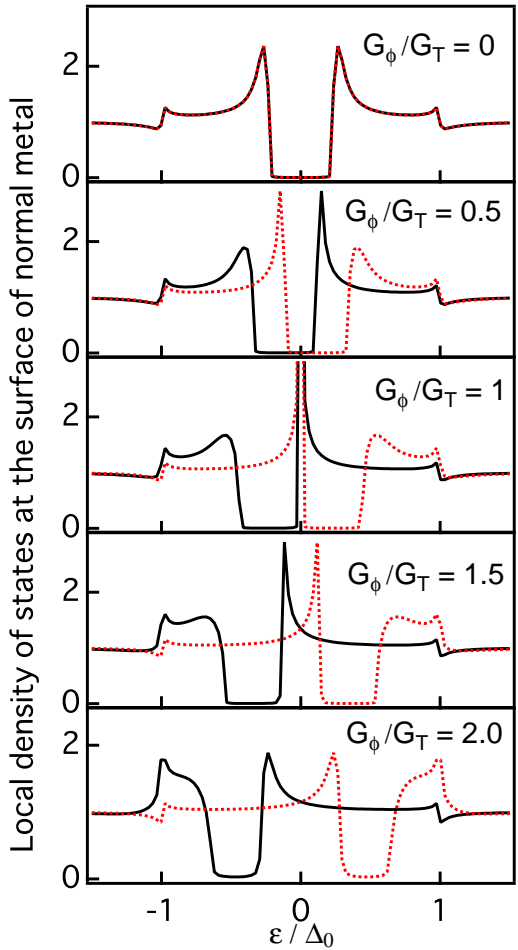}
\end{center}
\caption{(color online).
Local density of states at the surface of the normal metal with $R_d/R_b=0.2$.
The results for $\nu=1$ and $\nu=-1$ are shown with the solid and broken lines,
respectively.
}\label{fig2}
\end{figure}

We first show the local density of states (LDOS) at a surface of the normal metal
for several choices of $G_\phi/G_T$ in Fig.~\ref{fig2}.
Here we choose the parameter $R_d/R_b=0.2$, which means the proximity effect is weak.
\begin{figure}[ht]
\begin{center}
\includegraphics[width=8cm]{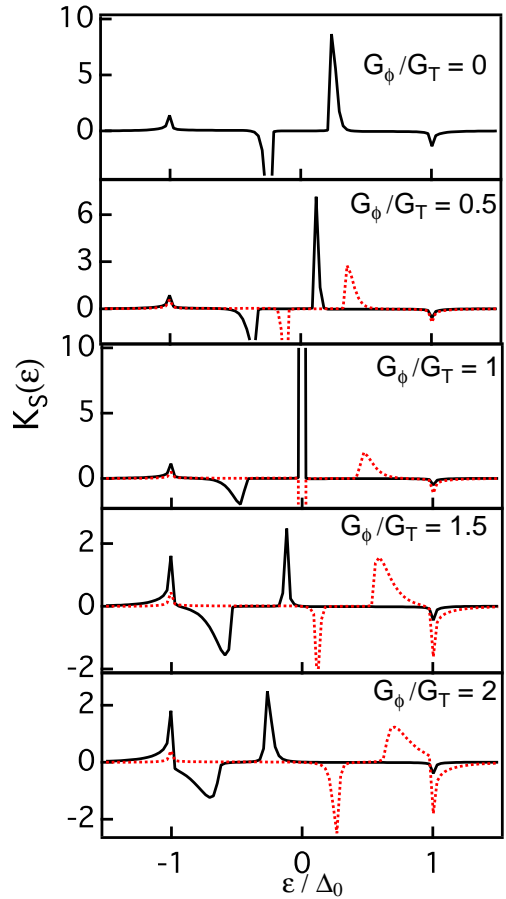}
\end{center}
\caption{(color online).
$K_s$ function at the surface of the normal metal.
The local pair density is calculated from Eq.~(\ref{ns}).
}\label{fig3}
\end{figure}
The solid and broken lines are the results calculated for $\nu=1$ and -1, respectively.
At $G_\phi/G_T=0$, two LDOS for $\nu=\pm 1$ are identical to each other and show
the minigap structure for $|\epsilon| \leq 0.2 \Delta_0$ due to the proximity effect.
In Fig.~\ref{fig3}, we show the pair spectral density defined in Eq.~(\ref{ks}).
At $G_\phi/G_T=0$, $K_s$ has a large positive peak around $\epsilon \approx 0.2\Delta_0$.
Therefore the local pair density in
Eq.~(\ref{ns}) becomes positive. This means the penetration of even-frequency pairs into
the normal metal.
When we introduce $G_\phi/G_T$ at 0.5, LDOS for $\nu=1$ shifts to negative direction, whereas
that for $\nu=-1$ moves to positive direction.
Correspondingly large positive peaks in $K_s$ are separated into two as shown
in Fig.~\ref{fig3}.
At $G_\phi/G_T=1.0$, two peaks in LDOS overlap each other. In $K_s$ function, the large positive
peak for $\nu=1$ totally cancels the large negative peak for $\nu=-1$.
As discussed in a previous paper~\cite{linder}, $G_\phi/G_T=1.0$ is a critical value.
For $G_\phi/G_T<1.0$, the even-frequency Cooper pairs is dominant in the normal metal.
On the other hand for $G_\phi/G_T>1.0$, the fraction of odd-frequency Cooper pair increases
with increasing $G_\phi/G_T$.
In particular at $\epsilon=0$, the frequency symmetry of Cooper pairs is purely odd.
When we increase $G_\phi/G_T=$ 1.5 and 2.0, the minigap in two subspaces are separated from
each other as shown in Fig.~\ref{fig2}. At the same time, $K_s$ has a large negative
peak in low energy region, which means the penetration of odd-frequency Cooper pairs into
the normal metal.

\subsection{Impedance}
\begin{figure}[ht]
\begin{center}
\includegraphics[width=9cm]{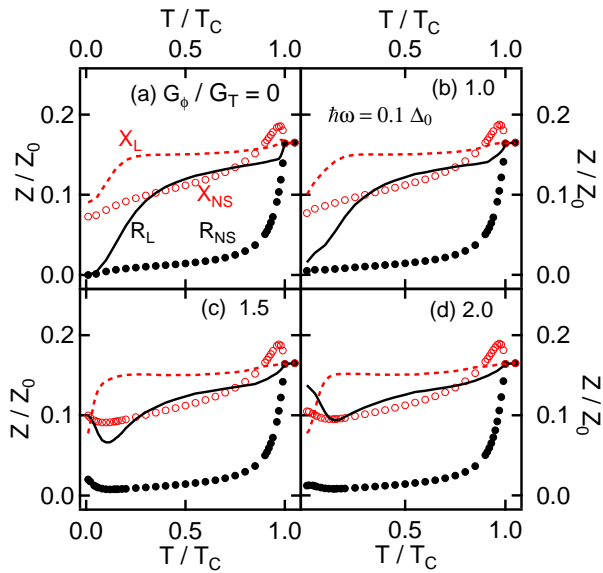}
\end{center}
\caption{(color online).
Impedance is plotted as a function of temperature for $R_d/R_b=0.2$ and
$\hbar\omega=0.1\Delta_0$.
The length of the normal metal is fixed at $L/\xi_{T_C}=1$.
The symbols ($R_{\textrm{NS}}$ and $X_{\textrm{NS}}$)
represent the results of impedance for the whole NS bilayer in Eq.~(\ref{zns1}).
The lines ($R_L$ and $X_L$) are the local impedance at
the surface of the normal metal given in Eq.~(\ref{zl}).
}\label{fig4}
\end{figure}

Next we show the impedance as a function of temperature
for several choices of $G_\phi/G_T$ in Fig.~\ref{fig4}, where
$\hbar\omega=0.1\Delta_0$.
We choose a boundary parameter as $R_d/R_b=0.2$ in Fig.~\ref{fig4},
which again means the proximity effect is weak.
The results for $G_\phi/G_T=0$ in Fig.~\ref{fig4}(a) shows the typical and conventional
behavior of impedance in NS junctions. The local impedance at the surface of normal metal
$R_{\textrm{L}}$ and $X_{\textrm{L}}$
monotonically decrease with decreasing temperature far below $T_C$
and satisfy the robust relation $R_{\textrm{L}} < X_{\textrm{L}}$.
The impedance of a NS bilayer $R_{\textrm{NS}}$ and $X_{\textrm{NS}}$ show qualitatively
similar behavior. Namely the impedance satisfies $R_{\textrm{NS}}<X_{\textrm{NS}}$.
These behavior are a direct consequence of the fact that
 all Cooper pairs belong to
even-frequency spin-singlet $s$-wave pairing symmetry.
Such characteristic feature remains even if we introduce
$G_\phi/G_T$ by small amount up to 1.0 as shown in Fig.~\ref{fig4}(b).
In the presence of the spin-dependent potential
at the NS interface, the odd-frequency spin-triplet $s$-wave Cooper pairs
appear in the normal metal
in addition to conventional even-frequency spin-singlet $s$-wave pairs.
 The fraction of odd-frequency pairs
is much smaller than that of even-frequency pairs for $G_\phi/G_T\leq 1$.
However $G_\phi/G_T$ exceeds unity as shown in Figs.~\ref{fig4}(c) and (d),
the the local impedance shows the unusual relation
$R_{\textrm{L}} > X_{\textrm{L}}$ at low temperature for
$T<T^\ast$, where $T^\ast$ is defined as the crossover temperature.
In Figs.~\ref{fig4}(c) and (d),
$T^\ast$ is about 0.02 for $G_\phi/G_T= 1.5$ and is 0.06 for $G_\phi/G_T=2$, respectively.
At the same time, $R_{\textrm{NS}}$ and $X_{\textrm{NS}}$ show the nonmonotonic
dependence of temperature for $T<T^\ast$.
For $G_\phi/G_T\geq 1$, the fraction of
the odd-frequency Cooper pairs become becomes larger than that of the even-frequency
pairs.
Thus the anomalous behavior of impedance in Figs.~\ref{fig4}(c) and (d)
is the direct evidence of the odd-frequency Cooper pairs in the normal
metal~\cite{ya11}.

\begin{figure}[ht]
\begin{center}
\includegraphics[width=9cm]{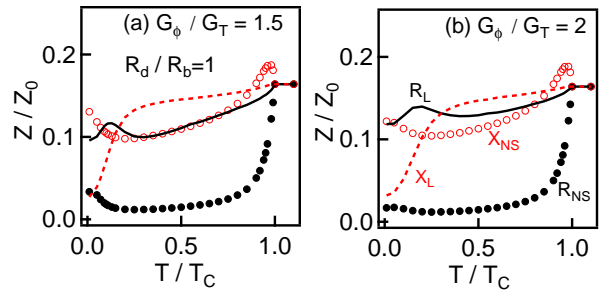}
\end{center}
\caption{(color online).
The results of impedance for $R_d/R_b=1$ and $\hbar\omega=0.1\Delta_0$.}
\label{fig5}
\end{figure}

\begin{figure}[ht]
\begin{center}
\includegraphics[width=9cm]{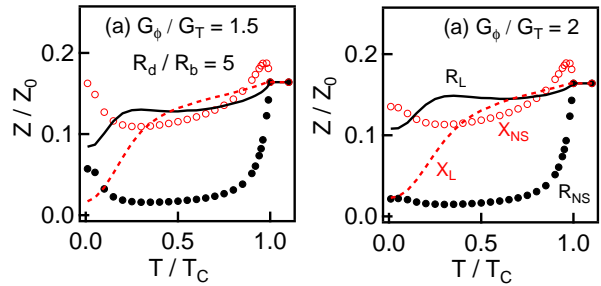}
\end{center}
\caption{(color online).
The results of impedance for $R_d/R_b=5$ and $\hbar\omega=0.1\Delta_0$.
}\label{fig6}
\end{figure}

Such anomalous behavior of impedance ($R_L>X_L$) is expected much wider temperature range
when we consider stronger proximity effect. The results are shown in Figs.~\ref{fig5}
and \ref{fig6},
where we choose $R_d/R_b=1$ in Fig.~\ref{fig5} and $R_d/R_b=5$ in Fig.\ref{fig6}.
The frequency of electromagnetic field remains unchanged from $\hbar\omega=0.1\Delta_0$
in both figures.
 At $G_\phi/G_T=1.5$, for instance, the crossover temperature
is $T^\ast=0.02T_C$ for $R_d/R_b=0.2$ in Fig.~\ref{fig4}(c),
 $T^\ast=0.17T_C$ for $R_d/R_b=1$ in Fig.~\ref{fig5}(a), and
$T^\ast=0.5T_C$ for $R_d/R_b=5$ in Fig.~\ref{fig6}(a).
In the same way at $G_\phi/G_T=2.0$,
$T^\ast/T_C$ is 0.06, 0.3, and 0.8 for in Fig.~\ref{fig4}(d),
 in Fig.~\ref{fig5}(b), and in Fig.~\ref{fig6}(b), respectively.
Thus we conclude that the anomalous relation in the local impedance
$R_L > X_L$ can be observed wider temperature range
for larger $R_d/R_b$.
On the other hand, the impedance of the whole NS bilayer always
shows the usual relation $R_{\textrm{NS}}<X_{\textrm{NS}}$.
The even-frequency Cooper pairs in the superconductor dominate the
impedance of the bilayer. The nonmonotonic temperature dependence of
$R_{\textrm{NS}}$ and $X_{\textrm{NS}}$, however, can be seen
for $T<T^\ast$.

\begin{figure}[ht]
\begin{center}
\includegraphics[width=9cm]{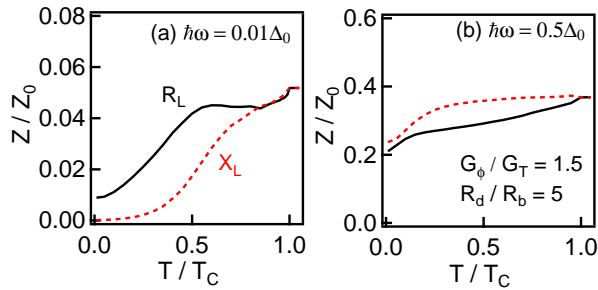}
\end{center}
\caption{(color online).
The results of local impedance for $R_d/R_b=5$ and  $G_\phi/G_T= 1.5$.
We choose $\hbar \omega/\Delta_0=0.01$ in (a) and 0.5 in (b). These results
should be compared with the results for $\hbar \omega/\Delta_0=0.1$ in Fig.\ref{fig6}(a).
}\label{fig7}
\end{figure}

Finally we look into the impedance for several choices of the frequency of
electromagnetic field in Fig.~\ref{fig7}, where we choose $G_\phi/G_T= 1.5$
and $R_d/R_b=5$. The frequency of electromagnetic field
is chosen as $\hbar\omega/\Delta_0=$ 0.01 and 0.5 in (a) and (b), respectively.
Here we focus only on the local impedance $Z_L$. These results should be compared
with Fig.~\ref{fig6}(a) for $\hbar\omega/\Delta_0=$ 0.1.
The crossover temperature to the anomalous relation is higher for
smaller frequency. In Fig.~\ref{fig7}(a), we find $T^\ast\sim 0.8T_C$.
Thus it is easier to detect the anomalous relation $R>X$ in lower frequency in
experiments. On the other hand, any sign for the odd-frequency pairs cannot be seen in
the results for high frequency at $\hbar\omega=0.5\Delta_0$
in Fig.~\ref{fig7}(b).
Thus we need to tune the frequency of electromagnetic field
to be much smaller than $\Delta_0/\hbar$.

On the basis of the calculated results, we predict that the anomalous relation
of the impedance $R>X$ due to the odd-frequency Cooper pairs would be observed
for high value of $G\phi/G_T$ and sufficiently low frequency of electromagnetic filed.
The fabrication of NS bilayers using a thin ferromagnetic insulator~\cite{blamire}
would realize large enough value of $G\phi/G_T$. 
It's also important to note that, as shown in Ref.~\cite{yoshizaki}, in the case of thin ferromagnetic (F) film
the effects of the exchange field and the $G\phi/G_T$ are equivalent. Therefore, the predicted anomalous
behavior of impedance can be also realized in S/F junctions with thin F-layer.
At the same time, the local impedance measurement is possible
now~\cite{machida}. Thus the conclusion of this paper could be confirmed in experiments.

\section{conclusion}
We have studied the surface impedance ($Z=R-iX$) of a normal-metal-superconductor bilayers
which has the spin-dependent potential at its junction interface.
The complex conductivity is calculated from the quasiclassical Green function
which is obtained by solving the Usadel equation numerically.
The Effects of the spin-dependent potential at the interface is considered
through the $G_\phi$-term in the Kupriyanov-Luckicev boundary condition at the junction
interface. The spin-dependent potential produces the odd-frequency Cooper pairs
in the normal metal.
We conclude that the local impedance in the normal metal
shows the unusual relationship $R>X$ when the odd-frequency Cooper pairs
become dominant in the normal metal. The predicted results can be observed
by recently developed local impedance measurement technique. 
In this paper, we consider spin-singlet $s$-wave superconductor as a bulk 
superconductor. 
It is a challenging issue to extend this calculation 
available for unconventional superconductor, 
spin-singlet $d$-wave \cite{d-wave}, spin-triplet $p$-wave \cite{p-wave}, and 
topological superconductors \cite{topological}. 
In these systems, it is known that 
Andreev bound state or Majorana fermion governs 
charge transport \cite{Review}.

\section{acknowledgement}
This work was supported by KAKENHI(No. 22540355) and
the "Topological Quantum Phenomena" (No. 22103002) Grant-in Aid for
Scientific Research on Innovative Areas from the Ministry of Education,
Culture, Sports, Science and Technology (MEXT) of Japan.

\end{document}